\documentclass{IEEEtran}
\usepackage{subfigure}

\usepackage[utf8]{inputenc}
\usepackage{amsmath,amssymb}

\usepackage{graphicx}
\usepackage{subfigure}
\usepackage{color}
\usepackage{array}

\usepackage{algorithm}
\usepackage{algorithmic}

\begin{document}

\title{Mitigation of Random Query String DoS via Gossip}

\author{\IEEEauthorblockN{Stefano Ferretti, Vittorio Ghini}\\
\IEEEauthorblockA{Department of Computer Science, University of Bologna\\
Bologna, Italy\\
\{sferrett, ghini\}@cs.unibo.it}
}

\date{}

\maketitle

\begin{abstract}
This paper presents a mitigation scheme to cope with the random query string Denial of Service (DoS) attack, which is based on a vulnerability of current Content Delivery Networks (CDNs). The attack exploits the fact that edge servers composing a CDN, receiving an HTTP request for a resource with an appended random query string never saw before, ask the origin server for a (novel) copy of the resource. Such characteristics can be employed to take an attack against the origin server by exploiting edge servers. Our strategy adopts a simple gossip protocol executed by edge servers to detect the attack. Based on such a detection, countermeasures can be taken to protect the origin server and the CDN against the attack. We provide simulation results that show the viability of our approach.
\end{abstract}


\section{Introduction}

In two recent papers, a Denial of Service (DoS) attack has been discussed that exploits a vulnerability of current Content Delivery Networks (CDNs) \cite{spectrum,original}. CDNs are commonly believed to offer their customers protection against application-level DoS attacks \cite{Poese:2010}. In fact, it is well known that, due to its vast resources, a CDN  can absorb typical DoS attacks without causing any noticeable effect to users. However, authors of \cite{original} have found an attack where the presence of a CDN actually amplifies the attack against a customer Web site. 

A CDN is composed of several \emph{edge servers} that are utilized to answer users' requests. Usually, a request to a Web site (\emph{origin server}) employing CDN technologies is invisibly routed to these other nodes that maintain replicated contents geographically distributed across the CDN. Upon a request routed to an edge server, if it does not have the content, which might be a large file, it retrieves it from the origin server where the Web site is hosted. Then, it passes that resource to the user. From that moment, the edge server maintains a copy of the resource; this way, subsequent requests for that content might be successfully completed without retrieving again that resource from the origin server. This operation mode allows to distribute the workload and protects the origin server from being swamped with requests.

According to \cite{original}, the basic problem is that based on the current implementation of CDNs, edge servers are not allowed to manage ``query strings''.
A query string is a string that is appended to the URL the client is targeting; these strings are usually employed to communicate parameters to the server during some HTTP request.
Now, since edge servers do not contain any logic related to the Web site, but they simply maintain replicated resources to distribute the load, when they receive some HTTP request with a random query string which is added to a URL, they treat such a request as new and pass it on to the origin server. 
The problem is that if the origin server is not expecting a query string, it removes it from the HTTP request and supplies the file. Summing up, if an attacker asks an edge server for a resource and appends to that request a random query string, the edge server will request such a resource to the origin server in turn, even if it already has it. For this request, the origin server sends such a resource to the edge server.

This way, an attacker can force an edge server to retrieve a copy of a large file from the origin server several times. Not only, it has been noticed that if the attacker cancels the connection immediately after requesting the resource, that resource transmission from the origin server to the edge server continues anyway.
A DoS attack can thus be implemented as follows \cite{original}. The attacker can retrieve a list of edge servers and send HTTP requests (with random query strings appended to such requests) to a large number of edge servers from a single machine. For each single request, the connection can be canceled after a while; hence, each single request requires little computing power. 

Such random query string DoS attack is directed towards the origin server, that spends a lot of its work and its bandwidth to send such resources to several, distinct edge servers.
It is shown that a single attack can have a long-lasting effect on the origin server. 

To cope with it, approaches such as data mining would at most enable to understand that an attack has been done to a server, ex-post.
Some mitigation schemes are outlined in \cite{original}, that nevertheless do not solve completely the problem. 
For instance, to protect against the random query string vulnerability, a content provider can setup its CDN service so that only URLs without query strings are accelerated by the CDN. However, this limits the flexibility of the CDN. In response to the identification of such a threat of CDNs, it seems that no modifications are going to be accomplished \cite{spectrum}.

To prevent the attacker from hiding behind a CDN, the edge server can pass
the client's IP address to the origin server any time it forwards a request to the
origin. This can be done by adding an optional HTTP header into the request.
Of course, the attacker can still attempt to hide by coming through its own intermediaries, such as a botnet, or public Web proxies \cite{Ager:2010,original}. 

In this work, we propose a simple strategy to face this attack. The idea is to resort to a simple gossip protocol among edge servers (and the origin server). Every time a request with a false query string is received by the origin server from an edge server, the origin server answers by sending the requested resource, as usual. However, it informs the edge server (via some additional information) that the query string was a false one. Of course, such information does not mean that the user is a malicious node, the request might be malformed for a number of other reasons. In any case, the edge server transmits an alert of an such erroneous request to other edge servers, via a gossip algorithm. This way, edge servers can become aware of a random query string DoS attack, if more edge servers notice that an high number of erroneous query string requests have been generated for a particular origin site. Upon detection of the attack, appropriate schemes may be adopted to solve the problem. For instance, edge servers can stop sending requests containing appended query strings to the origin server. We provide some simulation results that confirm that such a simple approach can be adopted to detect a random query string DoS attack, by just adding such a gossip algorithm between servers, without altering the basic behavior of the origin site and edge servers.

A final remark is related to the use of CDNs within clouds, and in general to the integration between these two worlds \cite{Broberg:2009,Chiu:2010}. These types of attacks may represent a possible threat for cloud technologies, where the allocation of the number of nodes (e.g.~edge servers) is optimized based on the traffic and the workload the service is subject to. Our solution can be viably exploited also within these kinds of architectures.

The remainder of the paper is organized as follows. Section \ref{sec:attack} outlines the random query string DoS attack. Section \ref{sec:alg} presents the approach proposed in this work to cope with it. Section \ref{sec:exp} describes the simulation scenario and obtained results.
Finally, Section \ref{sec:conc} provides some concluding remarks.

\section{Random Query String DoS Attack}
\label{sec:attack}

\begin{figure}[t]
   \centering
   \includegraphics[width=\linewidth]{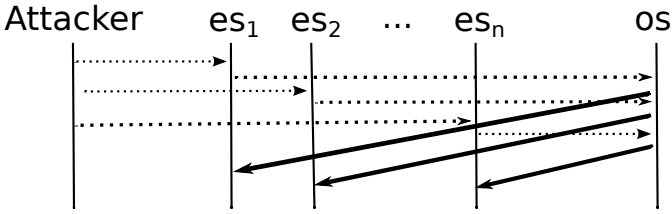}
   \caption{Random query string DoS attack.}
   \label{fig:attack}
\end{figure}

Figure \ref{fig:attack} shows how the random query string QoS attack works. For a detailed discussion the reader may refer to \cite{original}. First, the attacker needs to collect the addresses of edge servers $es_i$; there are several mechanisms to obtain their IP addresses \cite{Su:2006,original}. Then, it starts to make HTTP requests for some resources belonging to the origin server $os$ to edge servers; it appends random query strings to such requests, so that each $es_i$ will ask $os$ to provide it the resource. After a while, the attacker can cancel the HTTP request by closing the connection with $es_i$. For each received request, $os$ will send the requested resource to the corresponding $es_i$ nevertheless.

In the figure, requests from the attacker to the edge servers and requests from the edge servers to the origin server are depicted as horizontal lines, differently to resource transmissions from the origin server, to stress the fact that requests are lightweight, almost instantaneous messages, while file transmissions can take a while to be completed. This clearly wastes computational and communication resources of the origin server, and may cause a DoS.

\section{Coping with Random Query String DoS}
\label{sec:alg}

The target of a random query string DoS attack is the origin server. In fact, nodes in the CDN (edge servers) are exploited by the attacker to create a burst of requests towards it.
There are several problems concerned with mechanisms that simply try to detect such an attack at the origin server. 
For instance, one might try to determine the attacker by looking at the source of the request. However, the attacker may resort to mechanisms to vary the IP address, or it can hide behind some public proxy.
Another problem is that the attacker may change the file requested through edge servers; hence the origin server should look at all incoming requests. This implies an high computational load for the control.
Summing up, the origin server cannot do much by itself.

On the other hand, to tackle the problem it is probably better to avoid some complicated coordination scheme that involves all the edge servers for each request. In fact, this could easily slow down the edge servers responsiveness and strongly impact the effectiveness and the general performance of the CDN.

In this sense, the use of gossip algorithms could be of help \cite{simutools,disio11}.
Indeed, it has been recognized that gossip schemes can easily spread information through networks.
In this section, we propose a scheme that employs a gossip algorithm among edge servers to detect a random query string DoS attack.

\subsection{Overview of the Approach}

The scheme requires a simple extension at the origin and edge server and works as follows. 
Any time the origin server $os$ receives a request with a false query string from an edge server $es_i$  (as made during the attack), $os$  replies as usual by discarding the invalid query string and sending the resource.
But in addition, $os$ alerts $es_i$ that the query string was invalid. Such an additional information can be included as an option within the HTTP message containing the resource, or it might be included in a different message as well.

Upon reception of the alert from the origin server $os$, the edge server $es_i$ gossips it to other edge servers, including other alerts (if any) it received previously from $os$ or from other edge servers.
This allows edge servers to understand if more that an edge server has received a false query string directed to the same origin server $os$. If so, then maybe $os$ is under a random query string DoS attack.

It is worth mentioning that the reception of an erroneous query string does not implies that the origin server is under attack. Such kinds of requests can be received for a variety of reasons, including human errors and incorrect implementations of external mashups that exploit some kind of Web resources coming from the origin server. These external factors should not affect the behavior of the origin server and false positive detections must be avoided. Thus, the identification of a possible attack should happen only after a ``sufficient'' number of occurrences. Then, appropriate counter-measures can be employed such as, for instance, alerting (through a broadcast) all edge servers, which from that moment will process only HTTP requests without any appended query strings.

A central point of the approach is to quantify the ``sufficient'' number of alerts to suspect that an origin server is under a random query string DoS attack. Considering the percentage of erroneous requests over the total number of requests on a given time interval probably does not represent an appropriate choice, since such metric would take into consideration the popularity of the Web service hosted on the origin server.
Instead, we employ the following simple heuristics. Each edge server collects all the alert messages coming from the origin server or from the gossip protocol executed among edge servers in the CDN. This number is divided by the number $S$ of edge servers. When this value exceeds a given threshold, then a random query string DoS attack is suspected. Such a measure is an estimation of the number of erroneous query string received per edge server during a time interval $\Delta$. An erroneous query string is assumed to be a rare event; hence a non-negligible value of these received requests, when considered globally, for the whole CDN, may clearly indicate a possible attack.

\subsection{Gossip Algorithm}

\begin{algorithm}[t]
\caption{Gossip Protocol executed at $e_i$} \label{alg:simple}
\begin{algorithmic}
\STATE {\bf function \textsc{initialization}()}
\STATE $\mathit{v} \gets$ \textsc{chooseProbability()} %
\STATE

\STATE {\bf function \textsc{gossip}($os$)}%
\STATE \emph{msg} = collect all suspected activities towards $os$ during $\Delta$%
\FORALL{$es_j \in \text{CDN} \setminus \{e_i\}$} %
    \IF{\textsc{random()} $< \mathit{v}$}  %
        \STATE \textsc{send}(\emph{msg},$\mathit{es_j}$)%
    \ENDIF%
\ENDFOR
\STATE

\STATE {\bf main loop behavior}%
\STATE {\bf on} reception of an alert \OR timeout idle status
\STATE $os$ = select the origin server to control
\STATE \textsc{gossip}($os$)
\end{algorithmic}
\end{algorithm}

The gossip protocol is shown in Algorithm \ref{alg:simple} \cite{simutools}. It is a very simple push dissemination scheme that exploits a constant probability to spread information. The term ``push'' means that nodes decide to send information to other ones via independent and local decisions. Differently to pull based schemes, no direct requests are performed by receivers.
In substance, when an alert must be propagated, the edge server $es_i$ randomly selects the receivers using a probability value $\mathit{v} \leq 1$. In particular, each edge server $es_j, i\neq j$ is gossiped based on a probability determined by $\mathit{v}$. On average, the alert is thus propagated from $es_i$ to $\mathit{v} (S-1)$ edge servers, if $S$ is the number of edge servers in the CDN.

A DoS attack is accomplished during a limited time interval, since the goal is to overflow the origin server with a huge number of requests that should waste all the origin server's resources and saturate its network bandwidth. This claims for a rapid detection of a random query string DoS attack. For this reason, each edge server sends gossip messages to others not only after a reception of an alert from an origin server, but also periodically. The origin server $os$ to consider is determined based on the source of the received alert message (if any has been received), or randomly chosen among those for which an alert message has been received previously. Then, $es_i$ executes the \textsc{gossip}() procedure to disseminate information related to $os$. 

Another consequence, which is concerned with such a sudden spike in the requests to the origin server, is that the activity of edge servers can be monitored taking into consideration limited time intervals. For this reason, edge servers exchange suspected activities monitored during a moving time window $\Delta$. This reduces the amount of data to be managed, processed and exchanged among edge servers. Gossip messages are thus limited in size and the control procedure executed at edge servers requires limited computational efforts.

\section{Experimental Evaluation}
\label{sec:exp}

In this section, we report on a simulation we performed to assess whether the approach is able to detect random query string DoS attacks, when varying the configuration of a CDN, and to assess if the scheme is subject to false positives.

\subsection{Simulator Details}

We developed our own simulator to assess the proposed scheme. It was a discrete-event simulator written in C code; pseudo-random number generation was performed by employing the GNU Scientific Library \cite{gsl-web-2010}. The simulator allows to test the behavior of a given amount of edge servers for a settable number of time steps.
The attacker is simulated as a random process that sends random query strings to some of the edge servers during the simulation. Such a query is transformed into a resource request, and in turn into an alert generated by the origin server to the edge server. The simulator allows also to simulate non-malicious requests containing erroneous query strings towards some edge servers. Also these requests are generated by a random process (whose generation probability can be varied).

The behavior of edge servers in the CDN was implemented as detailed in the previous section. In particular, the simulator allows to vary all the parameters related to the protocol, such as the dissemination probability $v$ for gossiping messages, the threshold for suspecting that an origin server is under attack and the size of the time window employed to consider the aggregate of received alerts.

\begin{figure*}[t]
   \centering
   \includegraphics[angle=-90,width=.4\textwidth]{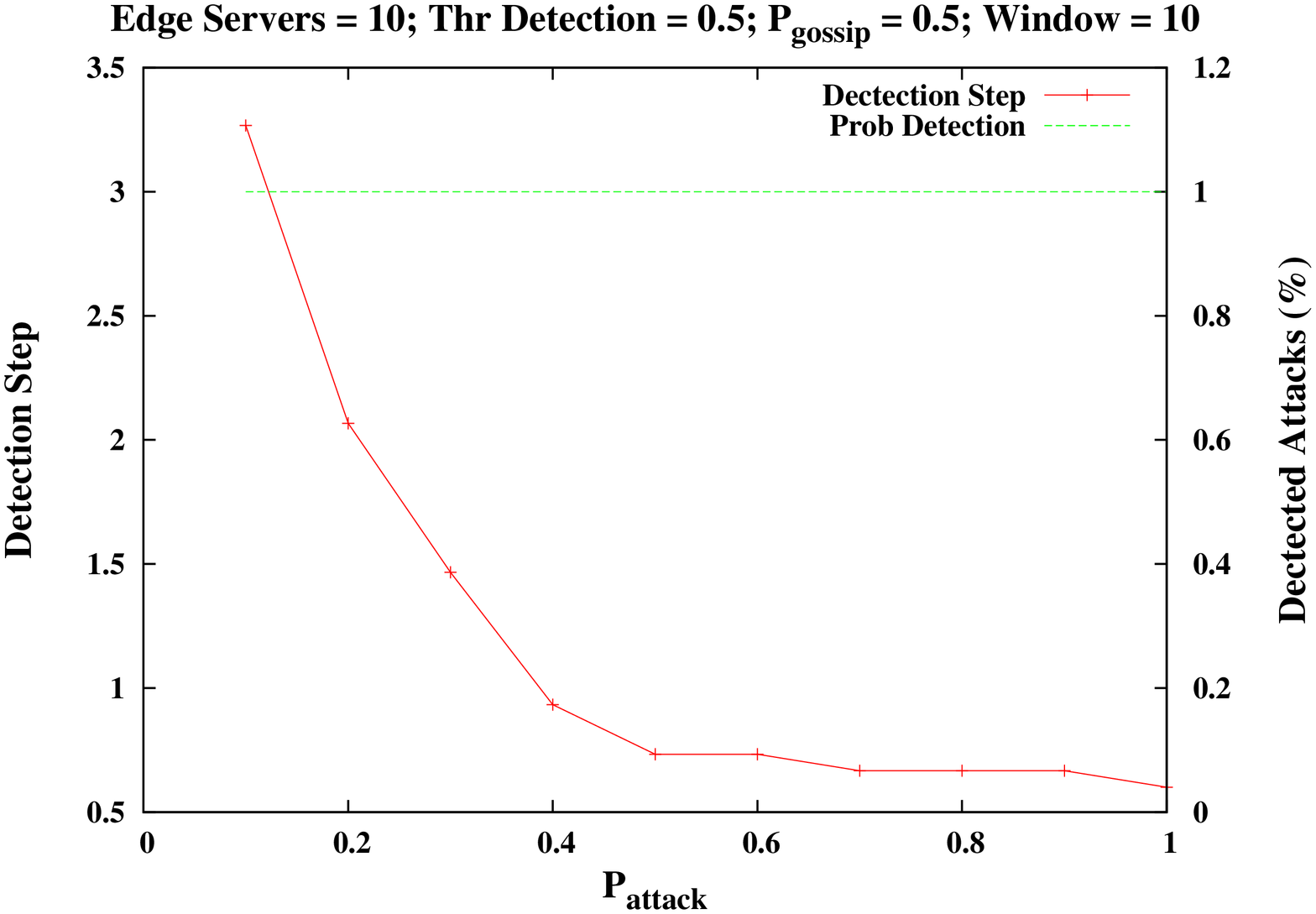}
   \includegraphics[angle=-90,width=.4\textwidth]{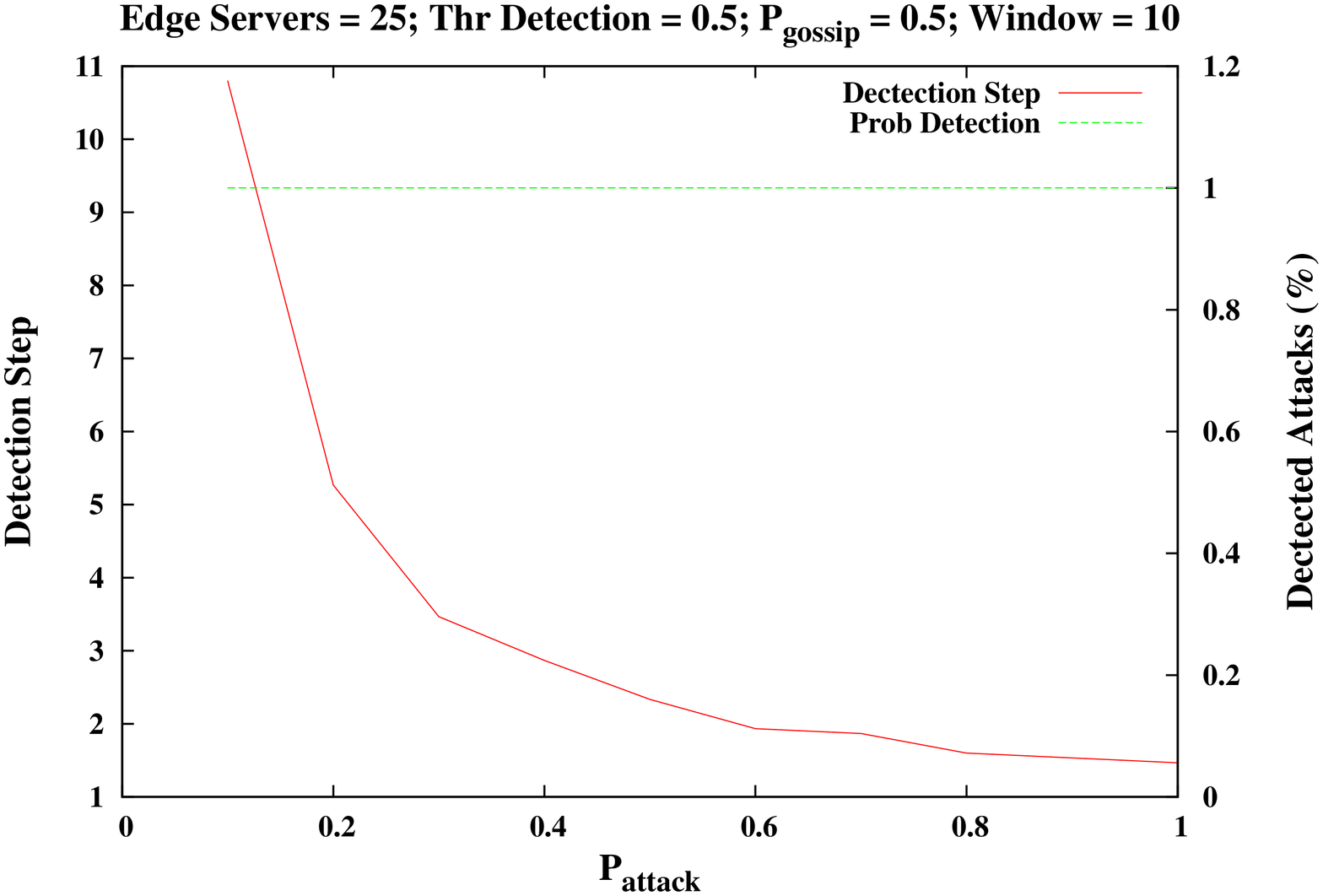}
   \includegraphics[angle=-90,width=.4\textwidth]{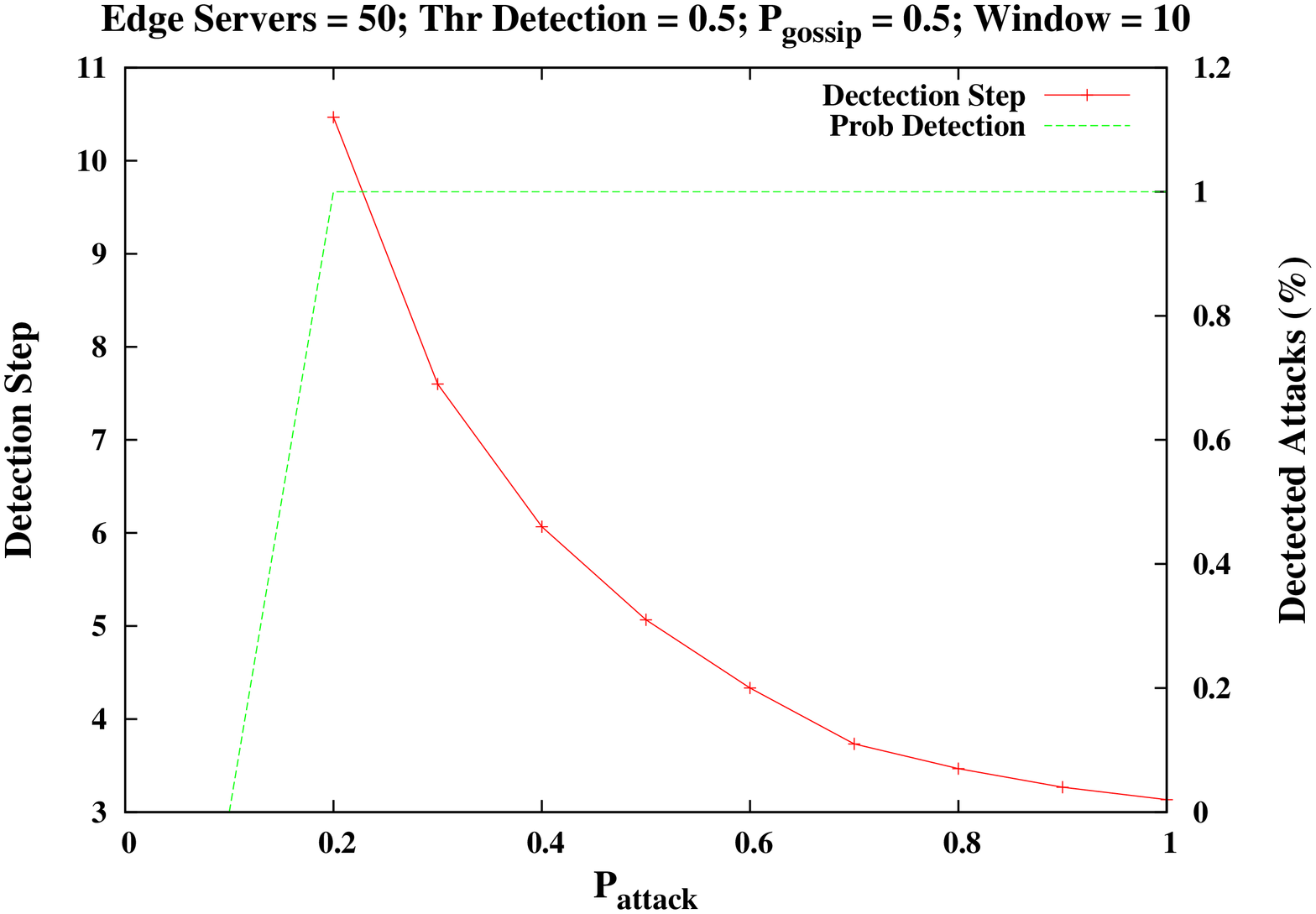}
   \includegraphics[angle=-90,width=.4\textwidth]{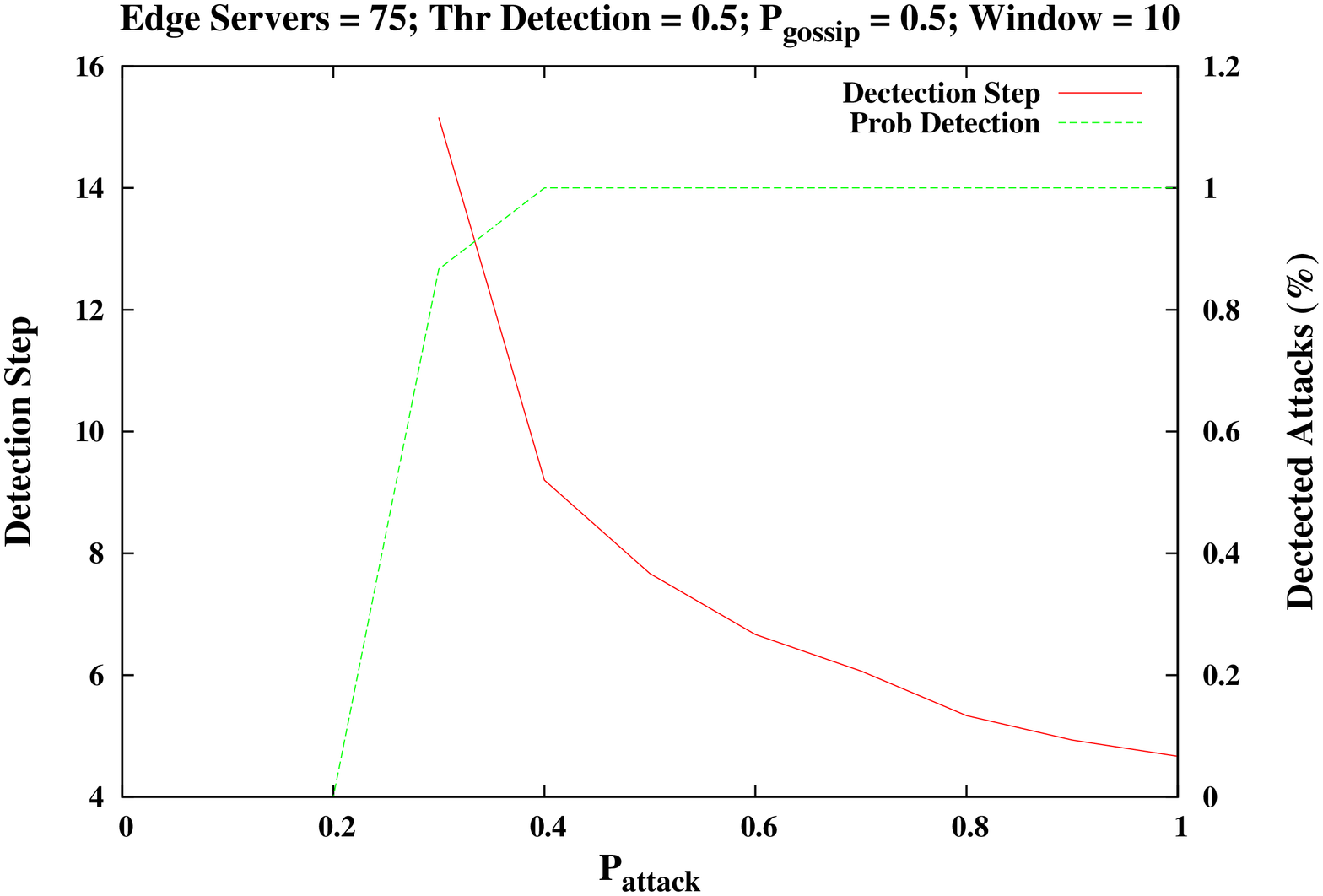}
   \caption{Average detection step and percentage of detection when varying the number of edge servers and the rate of generation of a random query string.}
   \label{fig:varia_es}
 \vspace{-0.2cm}
\end{figure*}

\subsection{Metrics of Interest and Configuration}

We performed a time-stepped simulation of duration $T=200$ steps. We varied the size of the moving time window.
We varied the probability of generation of a novel random query string request by the attacker to a given edge server $P_{attack}$ from $0.1$ up to $1$, while keeping the probability of a honest erroneous query string at a constant lower value (when not differently stated, its value is set equal to $0.01$).

In the following charts we show outlines when the probability of gossip $P_{gossip}$ among edge servers was set equal to $0.5$. We varied such value up to $0.9$ obtaining very similar results.
Another varied parameter is the size of the time window exploited to include alerts within gossip messages. We varied such value from $10$ up to $100$. Also in this case, we did not noticed significant differences. In the following, we show outcomes with a window size set equal to $10$ time steps (a lower value might have some impact on results).

A delicate aspect related to the success of the attack is concerned with the hardware configuration of the origin server, its computational capacity and network bandwidth, as well as the dimension of the resources requested by the edge servers to the origin server. Due to the extreme variability of these parameters, we decided to not exploit these metrics as those which determine if the attack succeeded. Rather, we exploited the already mentioned threshold to determine if the amount of received alerts at a given edge server enables to detect the attack. We varied the value set for such threshold from $0.25$ up to $1.5$. 
As discussed in the previous sections, we assume that when the system is not under attack, an erroneous query string is a rare event. The value compared against the threshold is an estimation of alerts received on average by each edge server during the considered time interval. Thus, given the typical number of edge servers $S$ in a CDN, the selected values represent non-negligible thresholds that might indicate an attack.

For each configuration setting, we run a corpus of 15 different simulations using different seed numbers. Results shown in the charts are obtained as the average of outcomes from the different simulation runs. The metrics we measured are mainly the number of steps needed by edge servers to detect that the origin server is under a random query string DoS attack, and the percentage of detected attacks.

\subsection{Results}

Figure \ref{fig:varia_es} shows the results of four different configuration scenarios in which we varied the number of edge servers in the CDN. In particular, the four charts refer to a configuration with $10, 25, 50, 75$ edge servers. Each chart reports the average step of detection after the beginning of the attack (red continuous line, y-axis on the left) and the percentage of detected attacks (green, dashed line, y-axis non the right). As shown in the charts, the higher $P_{attack}$  (i.e.~the stronger the attack to the origin server), the lower the number of steps required for detecting it and the higher the probability of detection.

It is clear that with an increased number of edge servers, more steps of interaction among these nodes is required to detect an attack. Moreover, in certain configurations the system is not able to detect all the attacks, when $P_{attack}$ has a low value, as shown in the figure.

\begin{figure*}[t]
   \centering
   \includegraphics[angle=-90,width=.4\textwidth]{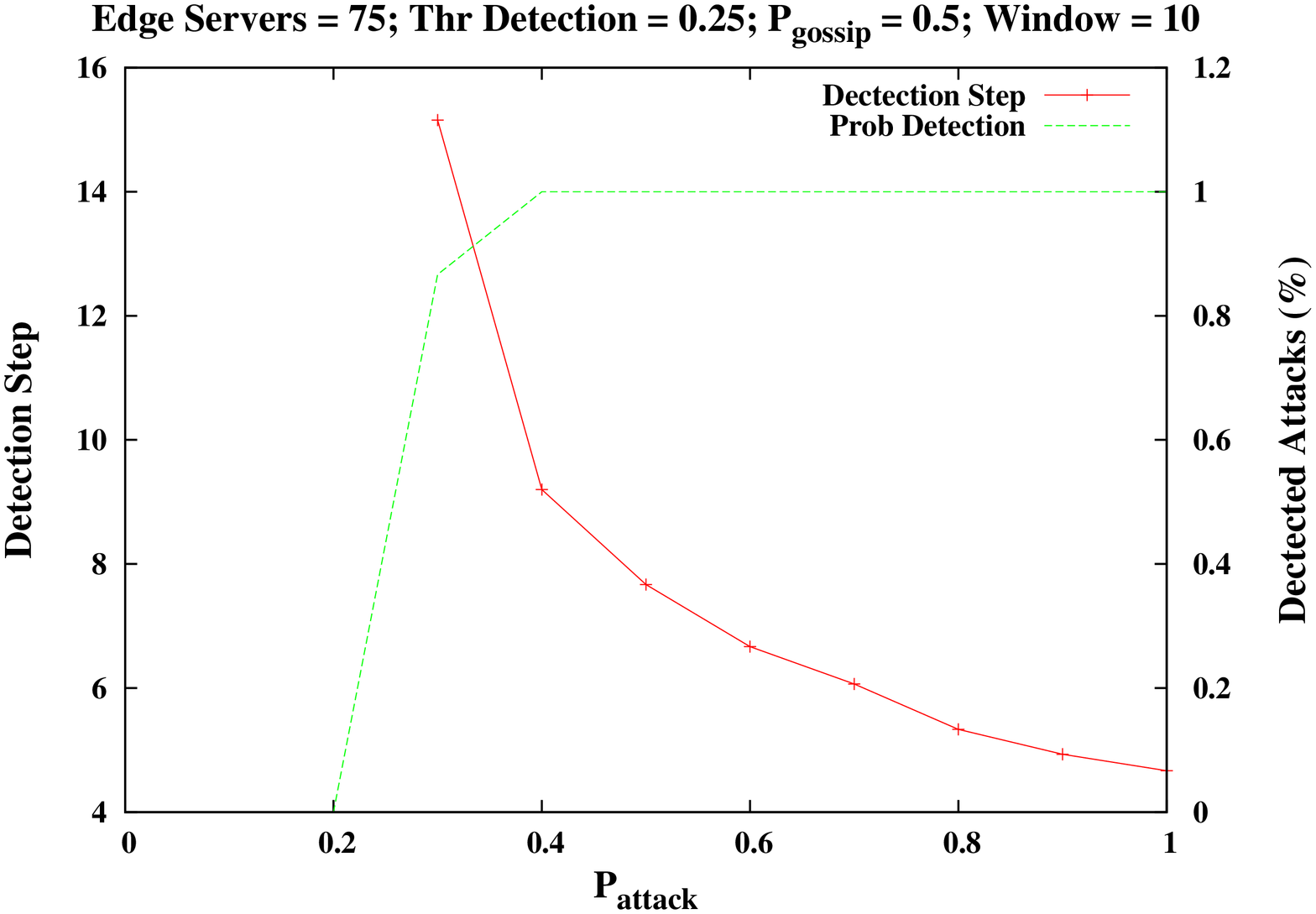}
   \includegraphics[angle=-90,width=.4\textwidth]{75_0.5_0.5_10_200_0.01_7.ps}
   \includegraphics[angle=-90,width=.4\textwidth]{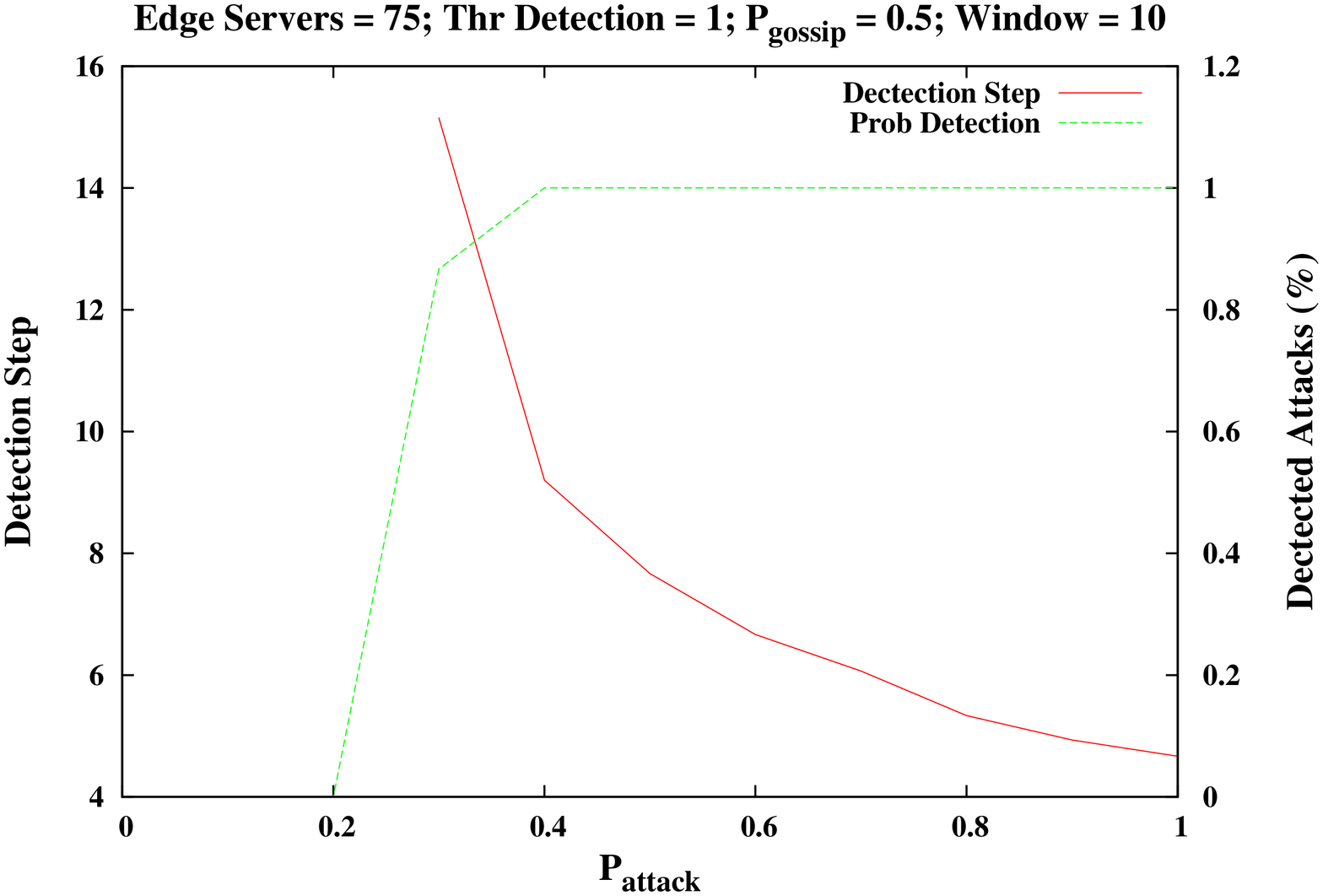}
   \includegraphics[angle=-90,width=.4\textwidth]{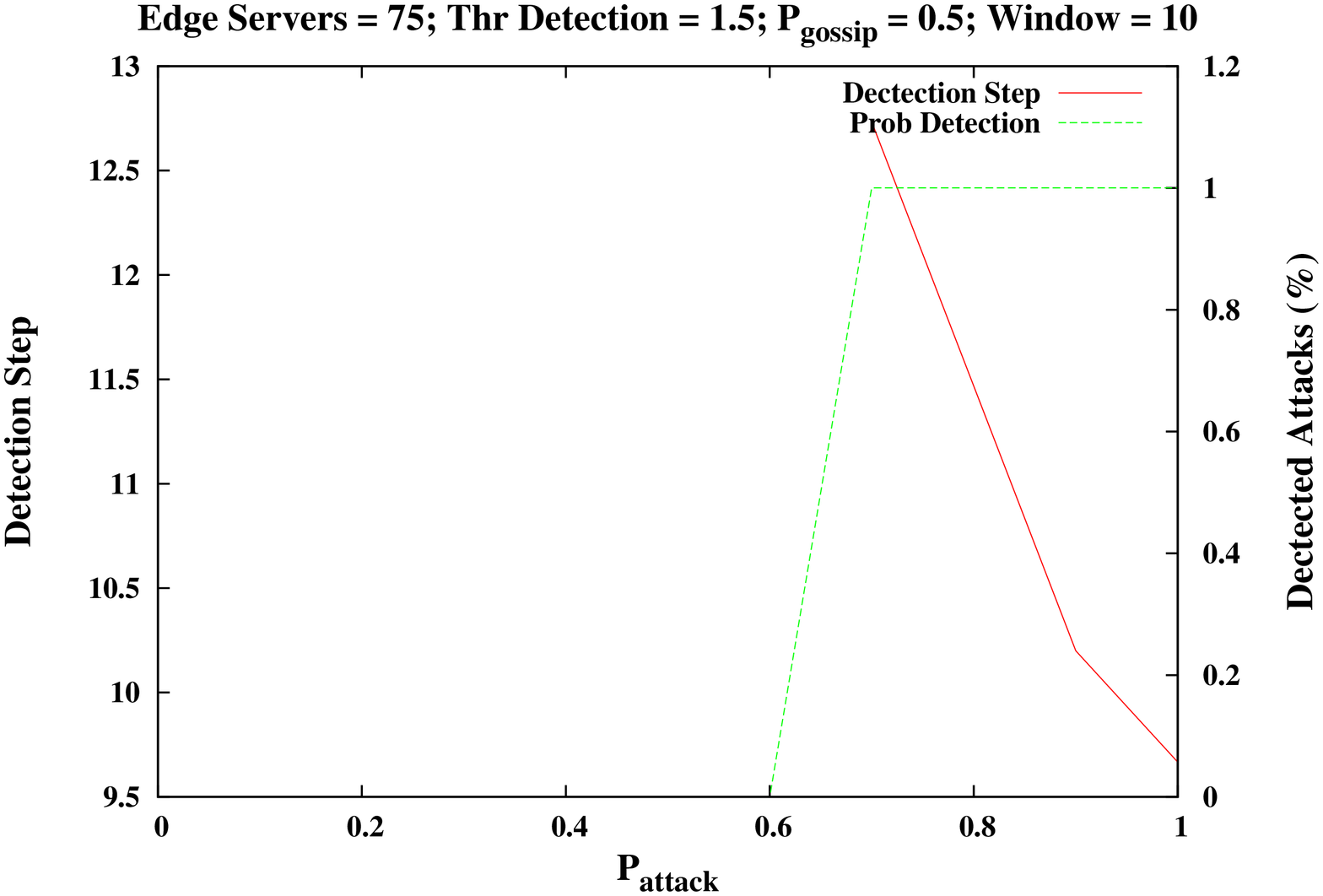}
   \caption{Average detection step and percentage of detection when varying the threshold to suspect a random query string DoS attack.}
   \label{fig:varia_thr}
 \vspace{-0.2cm}
\end{figure*}

Figure \ref{fig:varia_thr} shows the average detection step and percentage of detection obtained when the threshold employed to suspect a random query string DoS attack was varied from $0.25$ to $1.5$. In this cases, the number of edge servers was set equal to $75$ and the probability of gossip among edge servers was $0.5$. Charts show that, as expected, the tuning of this parameter influences the outcomes of the distributed scheme. In fact, we have very similar results when such parameter is kept below $1$; above it, results change and it becomes more difficult to detect an attack, mostly when a low probability of gossip $P_{gossip}$ is employed among edge servers.

It is worth mentioning that when we simulate the system not being under attack, but with possible generation of non-malicious erroneous query string requests, the system does not detect any DoS. In particular, we varied the rate of generation of such erroneous requests using a probability of a novel generation at each step, for each edge server, varying from $0.01$ up to $0.05$, with $P_{attack=0}$ and varying all other parameters as in the scenarios mentioned above. In this case, the CDN would behave normally.

\section{Conclusions}
\label{sec:conc}

This paper presented a scheme that may be effectively employed to mitigate random query string DoS attacks employed on CDNs. 
The idea is to exploit a gossip protocol executed by edge servers to detect if some origin server is under attack. The distributed scheme is simple and does not require particular efforts for the coordination among edge servers. 
Outcomes from simulations showed the viability of the proposed approach.

Further investigation may be devoted to understand if these kinds of DoS attacks can be led to cloud architectures, since these novel technologies may exploit CDNs (or similar solutions) to store and manage Web resources.

\end{document}